\documentclass[twocolumn]{aastex631}

\usepackage{times,graphics}
\usepackage{hyperref}

\usepackage{bookmark}
\usepackage{float}
\usepackage{graphicx}
\usepackage{dcolumn}
\usepackage{bm}
\usepackage{natbib}
\usepackage{pstricks}
\usepackage{epsfig}
\usepackage{epstopdf}
\usepackage{multirow}
\usepackage{amsmath}
\usepackage{lipsum}

\graphicspath{{Figures/}}

\newcommand{\be}{\begin{equation}}
\newcommand{\ee}{\end{equation}}
\newcommand{\bea}{\begin{eqnarray}}
\newcommand{\eea}{\end{eqnarray}}

\makeatletter
\newcommand*{\shifttext}[2]{%
	\settowidth{\@tempdima}{#2}%
	\makebox[\@tempdima]{\hspace*{#1}#2}%
}
\makeatother

\begin{document}



	\title{On the (higher multipoles) variance asymmetry in the cosmic microwave background}


	\author{MohammadHossein Jamshidi}
	\affiliation{Department of Physics, Shahid Beheshti University, 1983969411, Tehran, Iran}
	
	\author{Abdolali Banihashemi}
	\affiliation{Department of Physics, Shahid Beheshti University, 1983969411, Tehran, Iran}
	\affiliation{Department of Physics,	Sharif University of Technology, Tehran 11155-9161, Iran}

	\author{Nima Khosravi}
	\email{n-khosravi@sbu.ac.ir}
	\affiliation{Department of Physics, Shahid Beheshti University, 1983969411, Tehran, Iran}
	\affiliation{Department of Physics,	Sharif University of Technology, Tehran 11155-9161, Iran}

	\date{\today}

\begin{abstract}
We have studied the cosmic microwave background (CMB) map looking for features beyond cosmological isotropy. We began by tiling the CMB variance map (which are produced by different smoothing scales) with stripes of different sizes along the most prominent dipole direction. We were able to confirm previous findings regarding the significance of the dipole. Furthermore, we discovered that some of the higher multipoles exhibit significance comparable to the dipole which naturally depends on the smoothing scales. At the end, we discussed this result having an eye on  \textit{look-elsewhere-effect}. We believe our results may indicate an anomalous patch in the CMB sky that warrants further investigation.
\end{abstract}

\section{Introduction}
The observation of the cosmic microwave background radiation (CMB)  was a breakthrough in our understanding of the cosmos. Both WMAP \citep{wmap1} and Planck \citep{Planck:2018vyg} confirm the existence of an isotropic, homogeneous, Gaussian, nearly scale-invariant temperature fluctuations. Theoretically, the standard model of cosmology can describe the observations  accurately. This model is based on the cosmological principle\footnote{The isotropy and homogeneity of the large scales  is given by the very early inflationary era which is also responsible for the Gaussian and scale invariant fluctuation seeds.}, the standard model of particle physics, a period of (single field) inflation as well as the Einstein general relativity and includes the cosmological constant and cold dark matter ($\Lambda$CDM). 
	
Although, the successes of $\Lambda$CDM are  undeniable but there are some fractures in its triumph. In the theory side: i) we still do not know the foundations of dark matter and dark energy and ii) we face the famous cosmological constant problem \citep{Weinberg:1988cp}. In addition, there are reported observational tensions, between the CMB measurements \citep{Planck:2018vyg} and the local observations, in the values of: i) the Hubble constant \citep{Riess2022} and ii) the linear growth of structures \citep{Heymans:2020gsg}. Furthermore, there are some internal spatial anomalies in the  CMB e.g. i)  a dipole in the amplitude of CMB fluctuations and ii) the quadrupole-octopole alignment. They have been reported by WMAP \citep{deOliveira-Costa:2003utu} and then re-observed by Planck \citep{Schwarz:2015cma}. These kind of anomalies rise up questions about the cosmological principle and consequently the validity of $\Lambda$CDM, especially when they are considered altogether \citep{Muir:2018hjv,Hansen:2018pgg}. 

We focus on the dipole modulation anomaly which states that the amplitude of the temperature fluctuations in a certain hemisphere is higher than its value in its supplementary  hemisphere. This is a 2-3$\sigma$-ish anomaly which has been observed by WMAP \citep{eriksen2007hemispherical,eriksen2004asymmetries,hansen2009power,hoftuft2009increasing} and could not be removed from the CMB sky according to the Planck observation \citep{Planck:2019evm,quartin2015significance,flender2013small,axelsson2013directional,Akrami:2014eta}. The dipole modulation anomaly is against the statistically isotropic sky assumption and results in the existence of a preferred direction in the CMB sky.  

In this work, we would like to re-study this anomaly but in a more general scenario and looking for the higher multipoles modulation anomaly. From the data perspective, one should do it at least once to be sure that data does or does not show any new feature in higher multipoles. In addition from the theoretical perspective, realization of only dipole needs a fine tuning in the model building. In a previous work \citep{Banihashemi:2018has}, a dark energy model inspired by the physics of critical phenomena, we have proposed that this anomaly can be given because of an anomalous patch in the sky. This anomalous patch produces not only produces a dipole but also higher multipoles. Based on the above motivations, in the following we study the CMB fluctuations map and look for any higher multipoles signal.

\section{Data sets and methods}
\subsection{Map preparation}
In this analysis we have used \texttt{Commander} temperature map from \emph{Planck} 2018 data release. To assess the significance of the CMB signals, we have taken 1000 simulations into account with 300 noise maps available applied on them. Since the number of noises was not equal to the number of maps then each noise map has been employed 3 or 4 times. \emph{Planck}'s common mask is applied to all maps to remove the contaminated areas due to the foreground effects. We also degraded the mask such that if any degraded pixel contained an originally masked pixel, it was considered a mask.

\begin{figure}[h]
	\centering
	\includegraphics[width=0.7\linewidth]{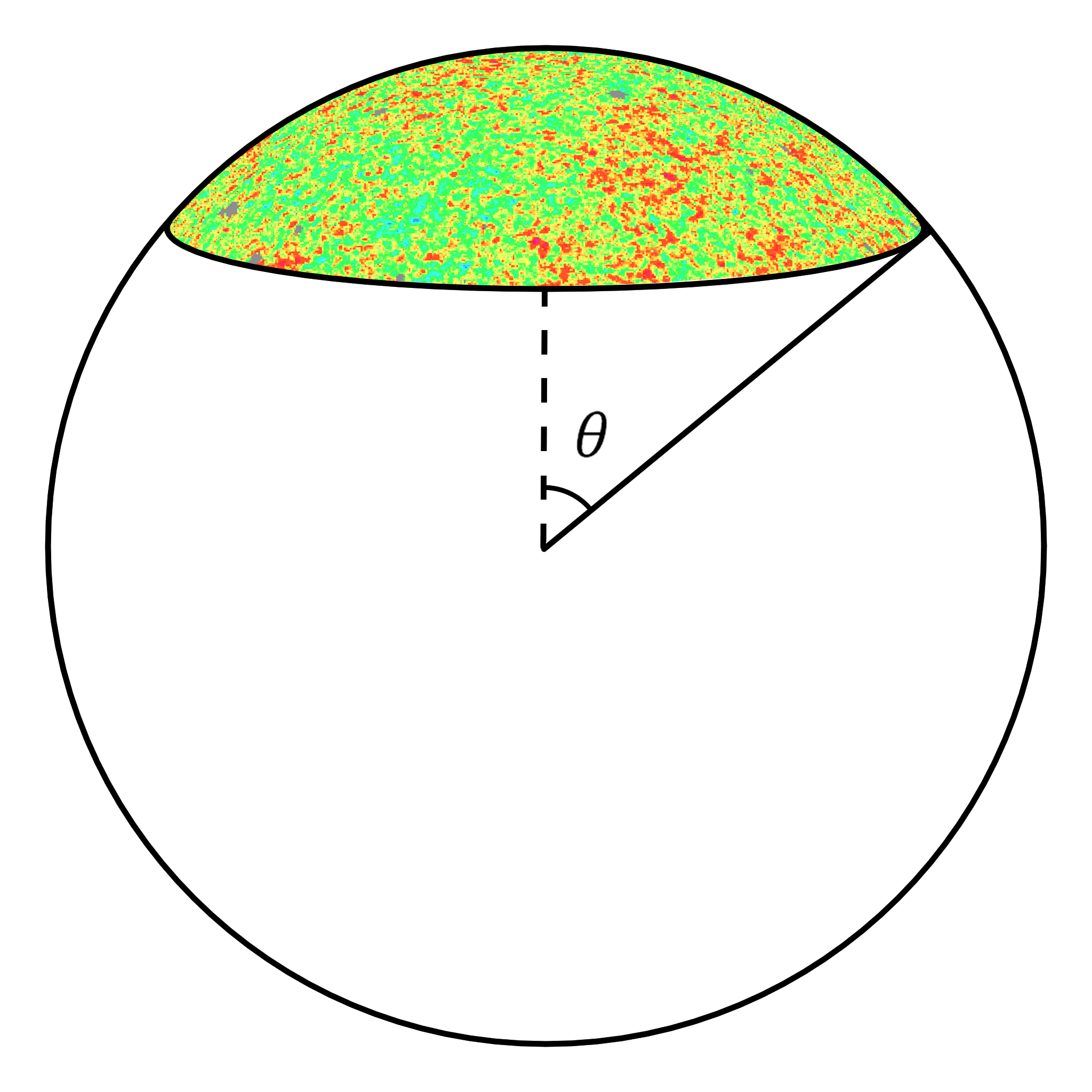}
	\caption{
		A typical cap (or disc) with radius $\theta$ whose variance corresponds to a pixel in the variance map.
	}
	\label{fig_stripe-1}
\end{figure}

The procedure of map preparation used here is pretty similar to the one used in \citep{Akrami:2014eta} \& \citep{ade2016planck}, for creating local variance maps, but with a newer data set. We have used the temperature maps discussed above, coarse-grained in HEALPix spherical subdivision \citep{Gorski:2004by} with $n_{\mathrm{side}} = 64$.
Then on these maps, we have considered 3072 caps (or discs; as a more familiar nomenclature in a flat-sky approximation) located on pixels of a HEALPix map with $n_{\mathrm{side}} = 16$, so that each pixel in this low resolution map is representative of a cap; see Figure \ref{fig_stripe-1}. 

The variance of each cap is then assigned to its corresponding pixel.
By this procedure, there also exist some caps, having considerable amount of masked pixels. Only caps with less than $90\%$ masked area, are included in the upcoming steps. Otherwise, we mark their corresponding pixels as masked in the local variance map.
Then, for each pixel in all maps (CMB and 1000 simulations), the mean value of the exact same location from all 1000 simulations is subtracted, and the resulting value is divided by the variance of the variance of these 1000 pixels.
So the \textit{weighted local variance} (WLV) value assigned to each pixel in this map would be:
\begin{eqnarray}
	\mathrm{pixel\;value} = \frac{\sigma^2_{\rm sample}(T) -\langle\sigma^2_{\mathrm{sim}}\rangle(T)} {\sigma^2(\sigma^2_{\mathrm{sim}})}\Bigg|_{\rm cap}.
\end{eqnarray}
Normalizing by the factor of ${\sigma^2(\sigma^2_{\mathrm{sim}})}$ gives less weight to more masked caps; because the more masked is a specific cap, the more variant its ensemble will be.

\begin{figure}[h!]
	\begin{center}
		\includegraphics[trim={1cm 1.2cm 1cm 0.6cm},clip , width=\linewidth]{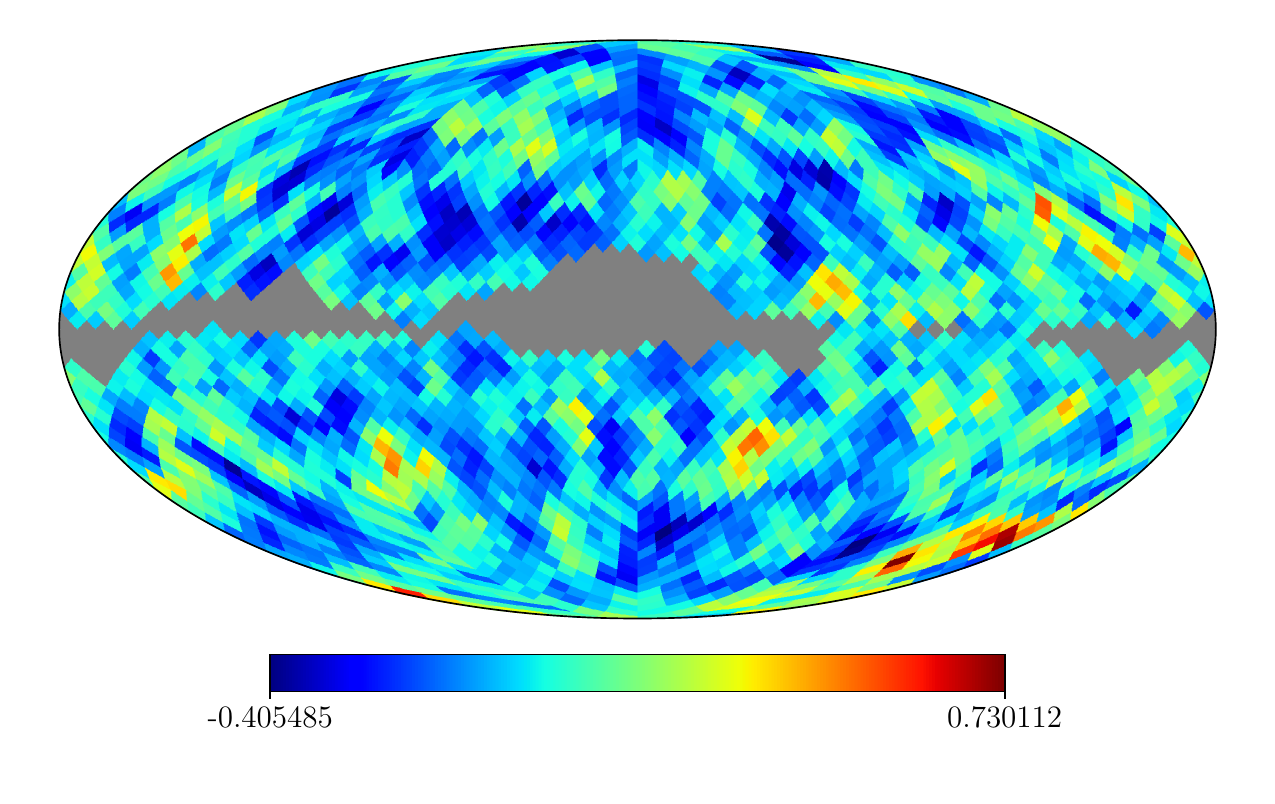}
		\includegraphics[trim={1cm 1.2cm 1cm 0.6cm},clip , width=\linewidth]{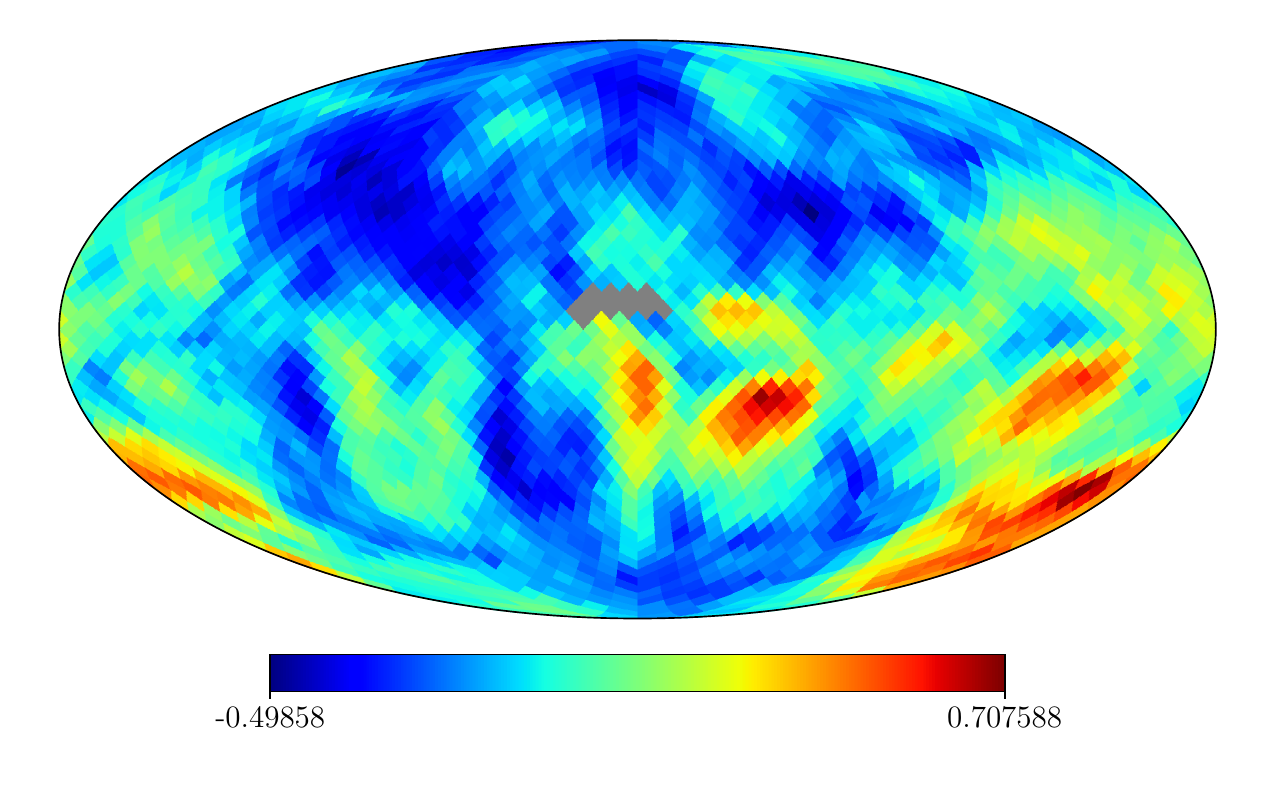}
		\includegraphics[trim={1cm 1.2cm 1cm 0.6cm},clip , width=\linewidth]{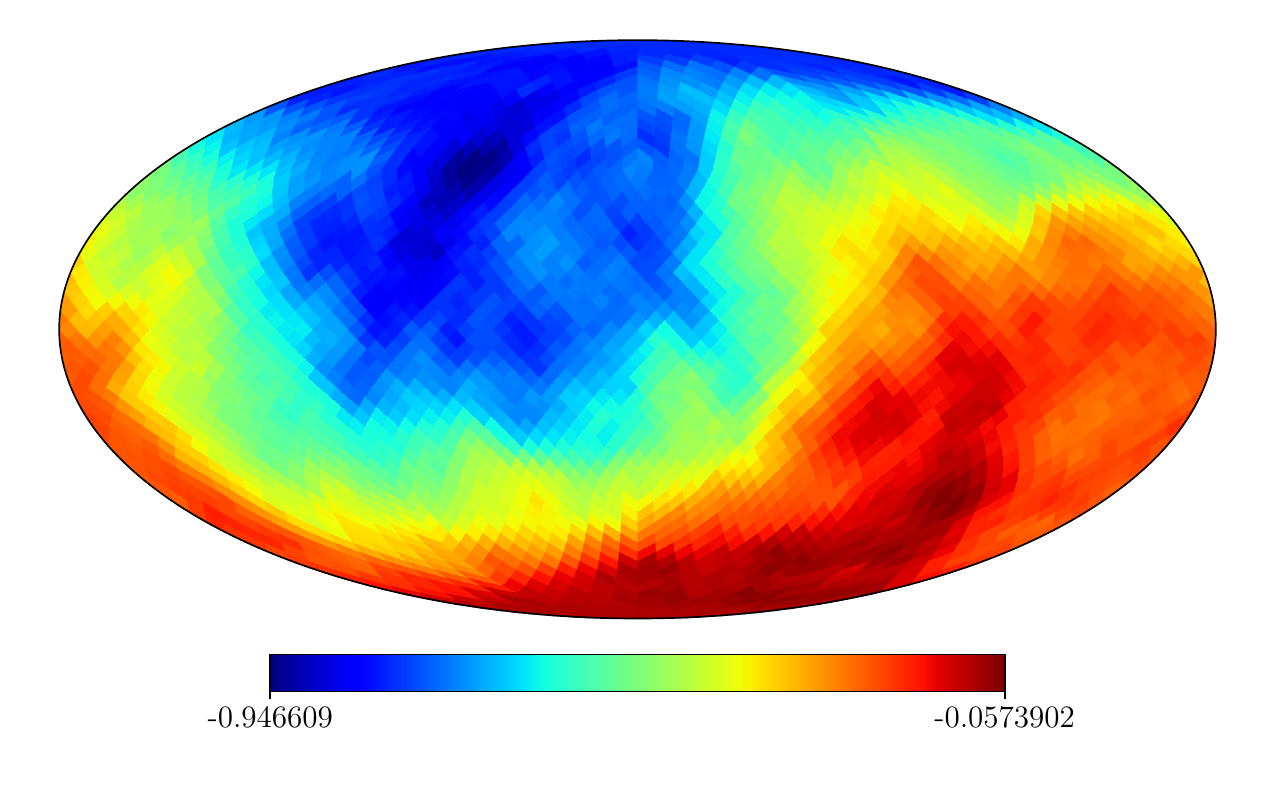}
		\caption{Weighted local variance maps created with caps of 7.0$^\circ$, 20.0$^\circ$ and 90.0$^\circ$ sizes. As these plots show, as the cap size increases, the higher $\ell$ details get washed out. For smaller cap sizes, it is clear that the map has more details than a simple dipole and the significance of it should be assessed. As for the masked area, note that our way of dealing with masked/unmasked regions is sensitive to the cap size; the bigger the cap size is, the more unmasked pixels we get.  Hence the masked region shrinks as the cap size increases.}
		\label{fig_local_var_maps}
	\end{center}
\end{figure}

It is also worth noting that by changing the radius of the caps, the local variance map and also the masked area will be changed due to different masked proportions in caps. We have prepared maps with different cap radii from $4^\circ$ to $90^\circ$, see Figure \ref{fig_local_var_maps}.

\subsection{expansion in legendre polynomials} \label{SubSec_Coefficients}
As a generalized application of the one dimensional Legendre expansion of $f(\theta)$ with coefficients:
\begin{eqnarray}
	A_l = \frac{2\ell+1}{2}\int f(\theta) P_\ell(\cos\theta)\, \sin\theta\, d\theta,
	\label{eq_legendre_expansion}
\end{eqnarray}
for a spherical function $f(\theta,\phi)$, one can replace the 1D integral with a normalized 2D integral and change the differential variable to the solid angle $\Omega$ like below: 
\begin{eqnarray}
	A_l = \frac{2\ell+1}{2} \times \frac{1}{2\pi}\int f(\theta, \phi) P_l(\cos\theta)\, d\Omega.
	\label{eq_generalized_legendre_expansion}
\end{eqnarray}
For the case of $f(\theta)$ this gives exactly the same result as the previous expression.
Now we define $\bar{f}(\theta)$ as the mean of those values of $f$ that are located on a ring(stripe) centered at polar angle $\theta$:
\begin{eqnarray}
	\bar{f}(\theta) = \frac{1}{2\pi}\int f(\theta, \phi) \,d\phi.
	\label{eq_stripe_mean}
\end{eqnarray}
Replacing this $\bar{f}(\theta)$ in the above equation gives the regular formula for the Legendre expansion coefficients.

\begin{figure}[h]
	\centering
	\includegraphics[width=0.7\linewidth]{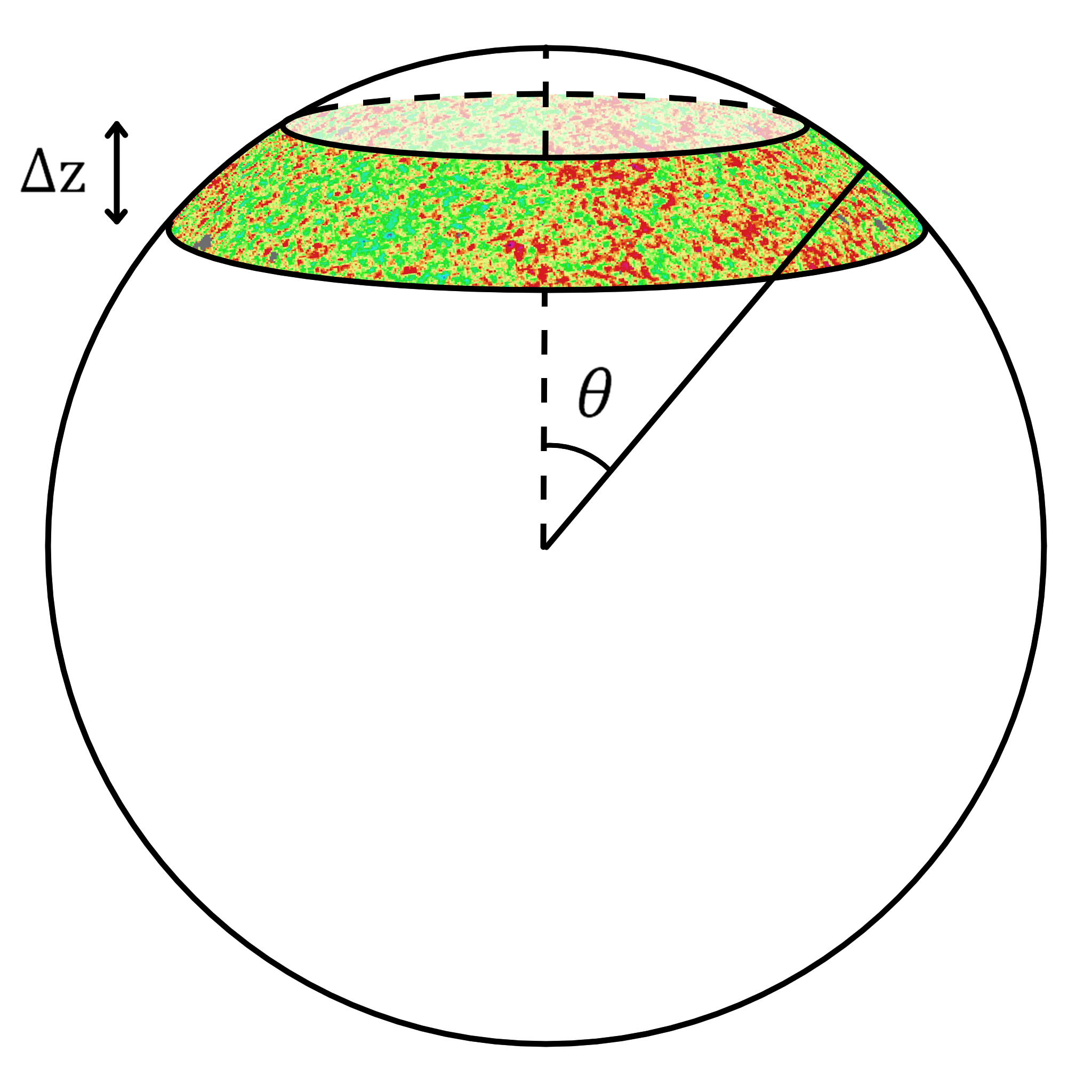}
	\caption{A typical stripe on the variance map that we take its average in order to create the function $f(\theta)$ used in Eq. \ref{eq_legendre_expansion}.}
	\label{fig_stripe}
\end{figure}

An important property of these coefficients is that the $A_\ell$ above is equivalent to the coefficients of the expansion of $f(\theta, \phi)$ with respect to $Y_{\ell m}$'s when $m=0$. If $m=0$, the (normalized) $Y_{\ell m}$ function simplifies to:
\begin{eqnarray}
	Y_{l0} = \sqrt{\frac{2\ell+1}{4\pi}}P_\ell.
	\label{eq_Ylm_and_Pl}
\end{eqnarray}
If we replace this in Eq. \ref{eq_generalized_legendre_expansion}, we get:
\begin{eqnarray}
	A_l = \sqrt{\frac{2\ell+1}{4\pi}}\int_\Omega f(\theta, \phi) Y_{\ell0}d\Omega.
	\label{eq_Al_from_Ylm}
\end{eqnarray}

Comparing to the regular coefficients of spherical harmonics expansion ($a_{lm}$), it would be clear that these $A_\ell$'s and $a_{\ell 0}$'s are simply proportional and differ up to a normalization factor:
\begin{eqnarray}
	A_l = \sqrt{\frac{2\ell+1}{4\pi}} a_{l0}.
	\label{eq_Al_and_alm}
\end{eqnarray}

\section{Dipole modulation revisited}
The well known dipole modulation in the literature \citep{Gordon:2005ai,Akrami:2014eta}, is defined in the form below:
\begin{eqnarray}
	f_{\mathrm{mod}}(\hat{n}) = (1 + A \,\hat{n}.\hat{p})f_{\mathrm{iso}},
\end{eqnarray}
in which $A$ is called the amplitude of the dipole and $\hat{p}$ the direction of it. The vector  $A\,\hat{p}$ is what typically reported as the dipole vector; for instance the \texttt{remove\_dipole} method of the \texttt{healpy} (HEALPix) package throws this vector as an output.

The term $\hat{n}.\hat{p}$ is nothing but the $\cos \theta$ and is equal to $P_1(\theta)$. Therefore, if we normalize a map to its monopole $f_{\mathrm{iso}}$, and compute Legendre dipole coefficient, we reasonably expect to get a similar amplitude to that of the dipole vector.

\subsection{Relation between coefficients and angular power spectrum}

In the literature \citep{Akrami:2014eta,ade2016planck} the angular power spectrum of the WLV is a base for infering the significance of the dipole and higher multipoles. This will result in the exclusive dominance of the dipole amplitude. In this section we aim to discuss the applicablity of the angular power spectrum for assessing the anomaly of WLV and the need for introducing a new way of the measuring the multipoles' amplitudes.

Hereafter, we use the word ``map" and ``function" interchangeably.
The angular power spectrum of a statistically isotropic function can be computed by the relation below:
\begin{eqnarray}
	\langle a_{\ell m} \,a^*_{\ell^\prime m^\prime}\rangle = C_\ell\, \delta_{\ell \ell^\prime}\, \delta_{m m^\prime}.
	\label{eq_cl_and_alm}
\end{eqnarray}
In the case of non-isotropy, the left hand side will also throw off-diagonal terms.

On the other hand, in Eq. \ref{eq_cl_and_alm}, $C_\ell$ is equivalent to the variance of $a_{lm}$'s with the same $\ell$; as for the isotropic maps, the mean of $a_{\ell m}$'s ($\langle a_{\ell m}\rangle$) is equal to zero. So the variance can be written this way:
\begin{eqnarray}
	\langle a_{\ell m} a^*_{\ell^\prime m^\prime}\rangle = \frac{1}{2\ell+1}\sum_{m} a_{\ell m} \,a^*_{\ell m}.
	\label{eq_alm_variance}
\end{eqnarray}
For the special case of $\ell=1$ the variance in the above expressions is equal to squared magnitude of the dipole moment of the map:
\begin{eqnarray}
	\langle a_{1m} a^*_{\ell^\prime m^\prime}\rangle = \frac{1}{3}(a^2_{1-1}+a^2_{10}+a^2_{11}).
	\label{eq_alm_variance_dipole_case}
\end{eqnarray}
If we rotate the map and align the $z$-axis to the dipole vector ($a_{1-1}, a_{10}, a_{11}$), and then again expand the function with respect to spherical harmonics for $\ell=1$, the only non-zero coefficient would be $a_{10}$ that is equal to the dipole amplitude of the map and $C_1$ is just $a^2_{10}/3$.
This property doesn't hold for any other $\ell$; and $C_\ell$ in other cases doesn't give the amplitude of its multipole but only results in the variance of $a_{\ell m}$ coefficients as discussed above.
Combining equations (\ref{eq_Al_and_alm}) and (\ref{eq_alm_variance_dipole_case}), one also finds a relation for Legendre coefficients: $C_1 = 4\pi A_1^2/9$.

An important point for the case of $\ell=1$ (Eq. \ref{eq_alm_variance_dipole_case}) is that since the rotation angles $(\theta, \phi)$ have two degrees of freedom, we can zero out two of the three coefficients in the Eq. \ref{eq_alm_variance_dipole_case} and have only one non-zero coefficient. For higher multipoles though, generally there is not such an option to keep only one coefficient left. Because the degrees of freedom for the coefficients are $2\ell+1$, and for the case of $\ell > 1$, this would be larger than $3$ degrees of freedom and removing two coefficients by rotation, generally leaves more than one non-zero coefficient. Therefore, the relation between $C_\ell$ and $A_\ell$ doesn't hold for $\ell > 1$.

As a final note, since the relation of $C_1$ and $a_{10}$ (and hence $A_{1}$) is monotonic, we expect the $p$-value of $C_1$ of any spherical map to be similar to that of dipole amplitude, but it is not correct for other multipole amplitudes.

\subsection{Numerical consistency of our dipole amplitude}
After aligning the map to the dipole direction, to expand the local variance map in terms of Legendre polynomials we proceed as follows: First, we should take the average of each ring of the map over azimuthal angle $\phi$ as described in Eq. \ref{eq_stripe_mean}, but since the map is discrete, we take a stripe with the width of $\delta z$ that covers areas that are $\delta z/2$ upper and lower than the ring located at polar angle $\theta$. Stripes prepared by this process, have equal areas and hence, contain similar pixel count. These stripes can overlap if the $\delta z$ is large, but it does not affect the final outcome, because stripes could be seen as window functions that somehow smoothen the function to some extent and can only wash out very small(high $\ell$) features that are not important to us. In the averaging procedure, we only consider pixels that are not masked. This whole process will result in a function that only depends on the polar angle $\theta$.

\begin{figure}[h]
	\begin{center}
		\includegraphics[trim={0.6cm 0cm 1.6cm 1cm},clip , width=\linewidth]{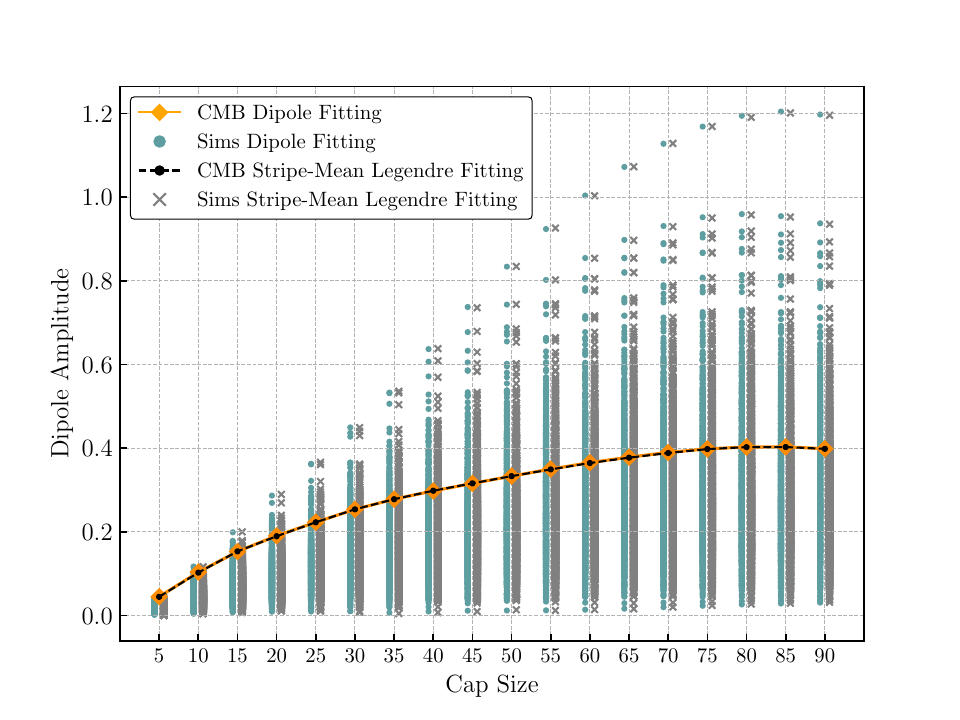}
		\caption{ This figure indicates the comparison of the dipole amplitude, between the "dipole fitting" and "stripe-mean Legendre fitting" methods. Here, the $\delta z$ of the stripe is set equal to the $\delta z$ of a cap with $10^\circ$ radius (1- cos($10^\circ$)). As expected, the dipole amplitudes taken from our method are in great agreement with the dipole fitting.}
		\label{fig_methods_comparison}
	\end{center}
\end{figure}

\begin{figure}[h!]
	\begin{center}
		\includegraphics[trim={0.4cm 0cm 1.2cm 0.8cm},clip , width=\linewidth]{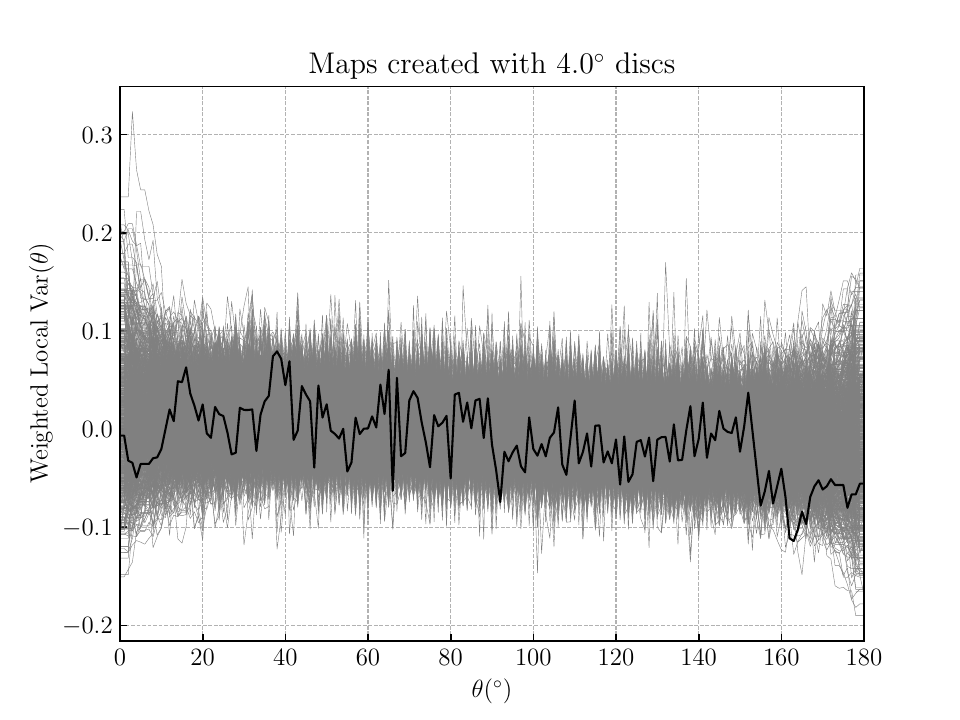}
		\includegraphics[trim={0.4cm 0cm 1.2cm 0.8cm},clip , width=\linewidth]{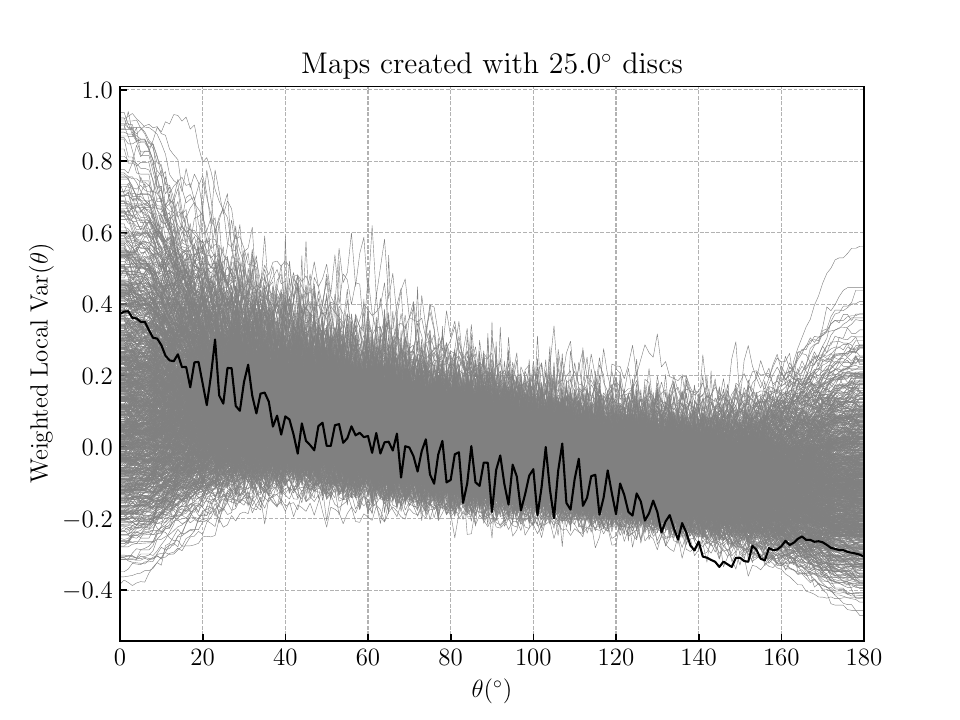}
		\includegraphics[trim={0.4cm 0cm 1.2cm 0.8cm},clip , width=\linewidth]{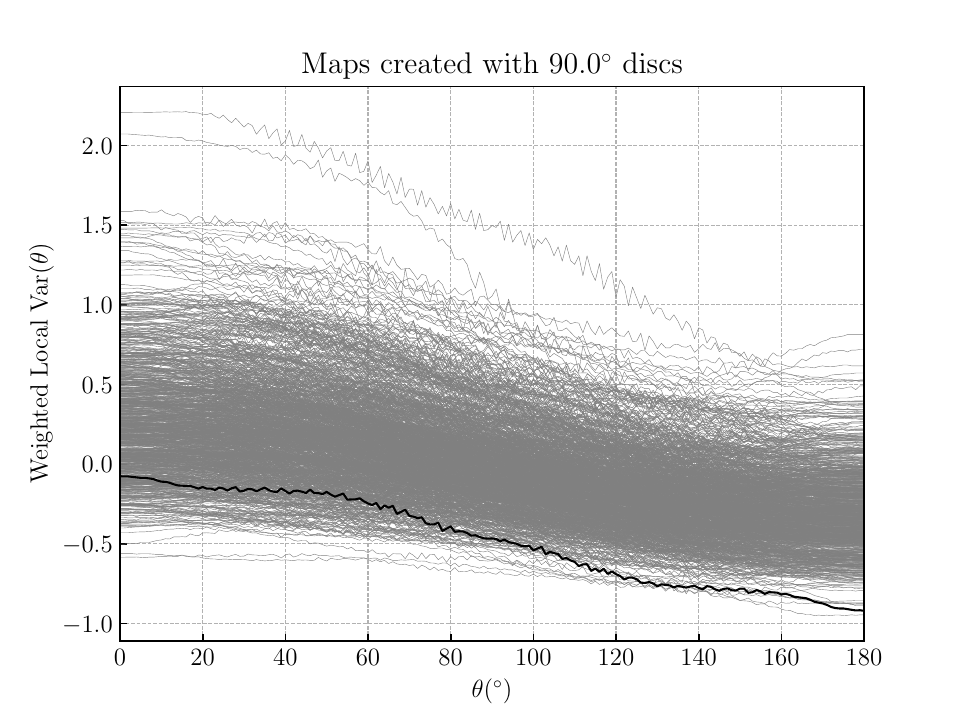}
		\caption{In these plots, we have shown the multipoles calculated by taking equal-area stripes and making a 1D function, the weighted local variance as a function of $\theta$. These plots show the dependence of the function on $\theta$ (the azimuthal angle). It is also seen that in some areas that are located roughly in the galactic north, the values are significantly lower than those of simulated skies.}
		\label{fig_local_var_of_theta}
	\end{center}
\end{figure}

We then use Eq. \ref{eq_legendre_expansion} to expand the 1D-function above in terms of the Legendre polynomials.
As seen in Figure \ref{fig_methods_comparison}, the resulting dipole amplitude from this method is very similar to that of dipole fitting, but this method also throws higher multipoles of the map, that we are interested in.

\section{Results}
\subsection{Spatial dependence of local variance}

The weighted local variance map clearly shows a dependence on direction (Figure \ref{fig_local_var_maps}), but the point is how significant these local variances are with respect to simulations. When comparing the values, whether in different patches of sky or different data series, one thing to note is having the mask on the maps. Since we align all maps to their dipole direction, the mask also rotates, and patches with similar relative locations will have different masked areas. By making use of stripes and creating a 1D function with the method explained in section \ref{SubSec_Coefficients}, we can overcome this issue if all the stripes contain more than $10\%$ unmasked pixels, just like the way the local variance itself is measured. Fortunately, this happens for 1000 \texttt{Commander} simulations, and we can compare stripe values. If that hadn't happened, since the simulation maps are created with the assumption of isotropy, we could do a pixel-wise map rotation (keeping the mask fixed) in a way that all the stripes cover a sufficient amount of unmasked pixels. If CMB itself had this issue, we could play with the stripe width ($\delta z$) to achieve the 1D function mentioned above. Nevertheless, it is possible to create the 1D function, from both CMB and simulations. Here we have taken the $\delta z$ to be similar to that of a cap with $10^\circ$ radius. The 1D functions for some local variance maps(with different disc sizes) are brought in Figure \ref{fig_local_var_of_theta}. It could be seen that there are areas of CMB sky located at the opposite direction of the dipole $(l,b)$ that have significantly lower local variances with respect to simulations. A viable cosmological model explaining such a feature on the CMB sky is beyond the scope of this work and needs its own dedicated effort.

\subsection{Higher multipoles}
\begin{figure*}[h!]
	\begin{center}
		\includegraphics[trim={1.6cm 0.2cm 4cm 1.2cm},clip , width=\linewidth]{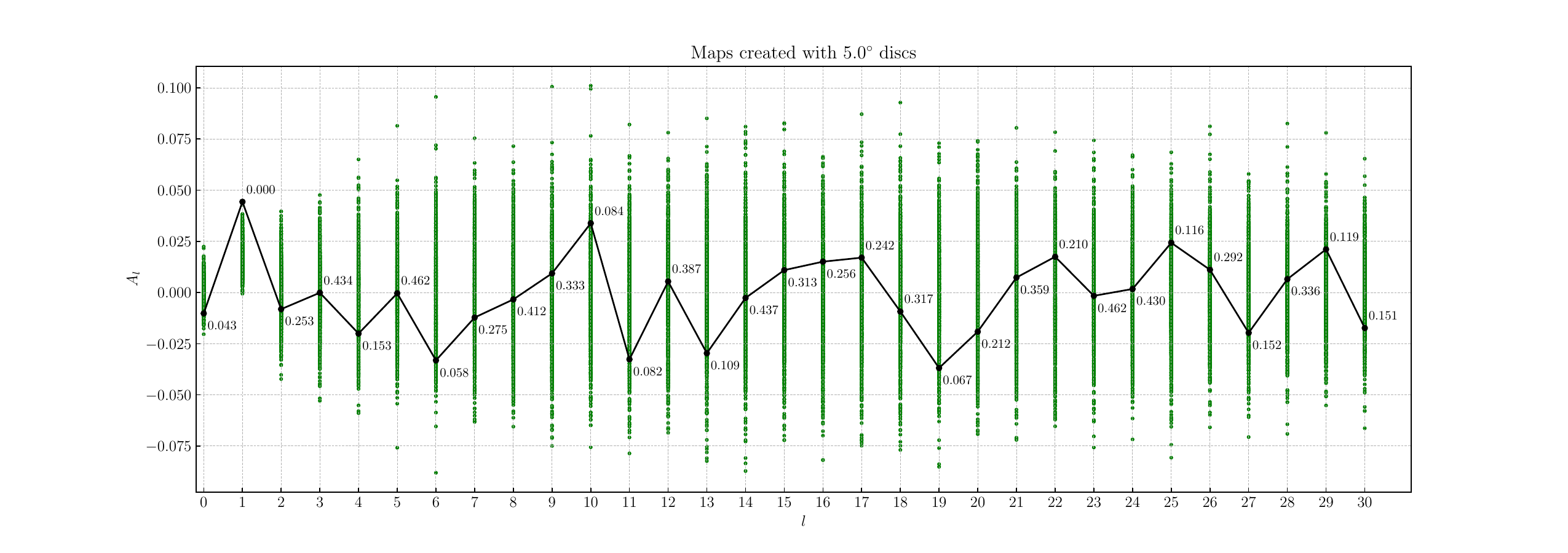}
		\includegraphics[trim={1.6cm 0.2cm 4cm 1.2cm},clip , width=\linewidth]{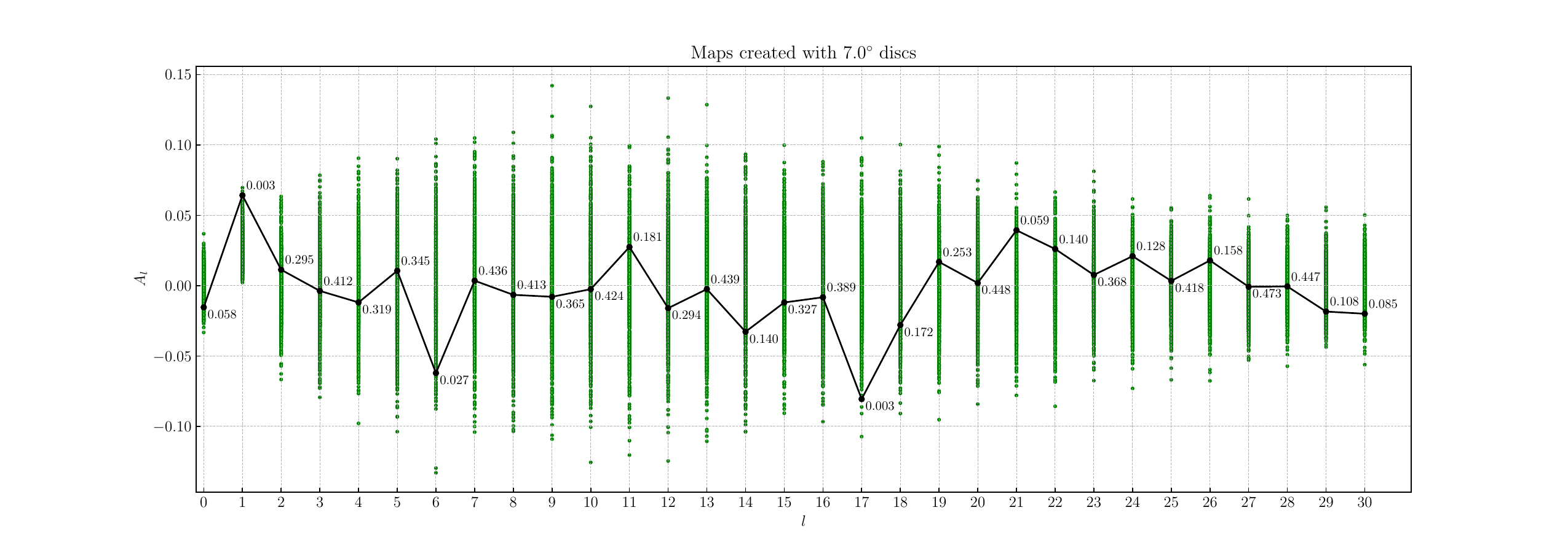}
		\includegraphics[trim={1.6cm 0.2cm 4cm 1.2cm},clip , width=\linewidth]{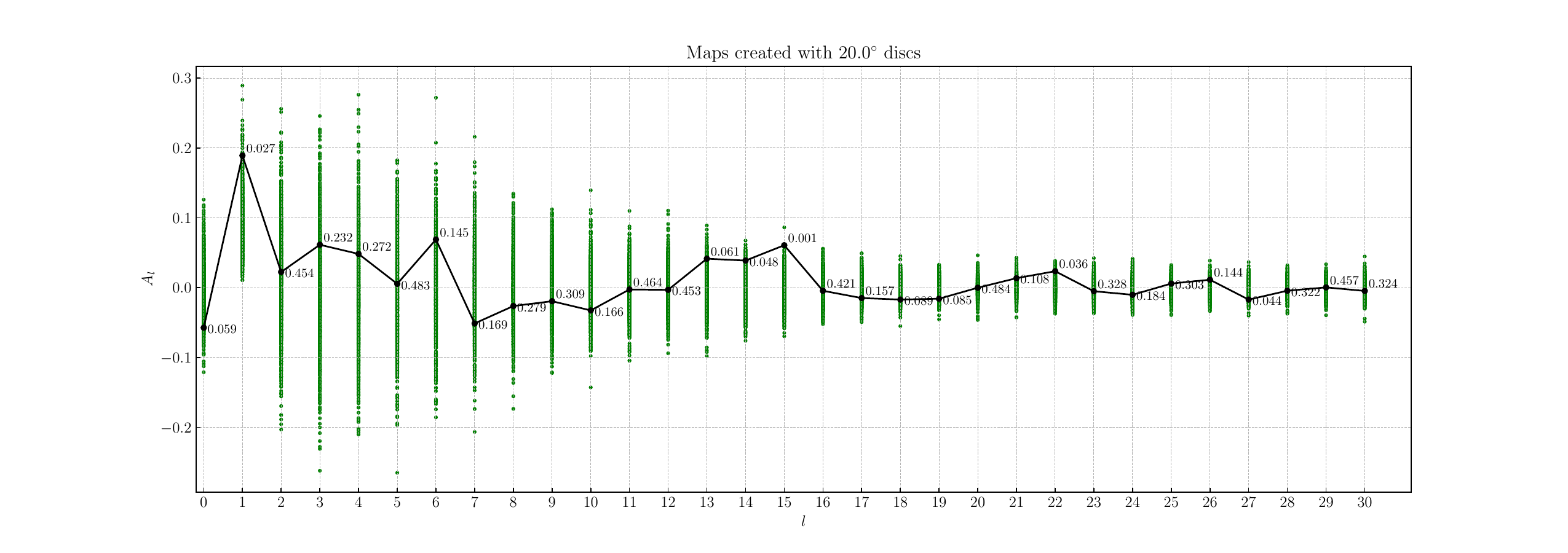}
		\caption{In these plots, the $WeightedLocalVar(\theta)$ is expanded in terms of Legendre polynomials, and their corresponding amplitudes are shown. The dipole amplitude is significant as it was previously reported in the literature, but here it is shown that there are al other multipoles that are as significant or even more significant than the dipole amplitude. For the map of $7^\circ$ discs, the amplitude of the multipole of $\ell=17$ has a $p$-value of 0.003, just like the dipole. For the map of $7^\circ$ discs, the one for $\ell=15$ has a $p$-value of 0.001 and it is more significant than the dipole amplitude which has a $p$-value of 0.027.}
		\label{fig_local_var_multipoles}
	\end{center}
\end{figure*}

As discussed in the last section and as seen in Figure \ref{fig_local_var_maps}, the local variance maps have finer structures, but the question is how significant these details are. To quantify this, by expanding the 1D function by Legendre polynomials (Eq. \ref{eq_legendre_expansion}), we can compare the multipoles taken from CMB data with the ones from simulations. We have done this procedure for different maps created with different discs of sizes from $4^\circ$ to $90^\circ$ to see if there are any significantly higher multipoles. In Figure \ref{fig_local_var_multipoles} we have shown the multipoles for three sample maps (disc sizes of $4^\circ$, $7^\circ$, and $20^\circ$). The dipole moment is significant for all these cases, but for instance, for the maps created with $7^\circ$ discs, the amplitude of the multipole of $\ell=17$ is as significant as the dipole, and both of them have a $p$-value of $0.003$. The situation gets even worse for the map of $20^\circ$ discs. The dipole moment has a $p$-value of $0.027$, whereas the amplitude of the multipole of $\ell=15$ has a $p$-value of $0.001$. It is worth noting that different disc sizes will smooth the map and make some local extrema wider or narrower, so the relevant $\ell$ that captures that specific bump size of the map would be changed and also shifted for different sizes. The existence of these significantly higher multipoles shows that the signal has more important details than a simple dipole that should be assessed, whether it is generated from the early universe or a late-time effect.

\begin{figure}
\begin{center}
	\includegraphics[width=9cm]{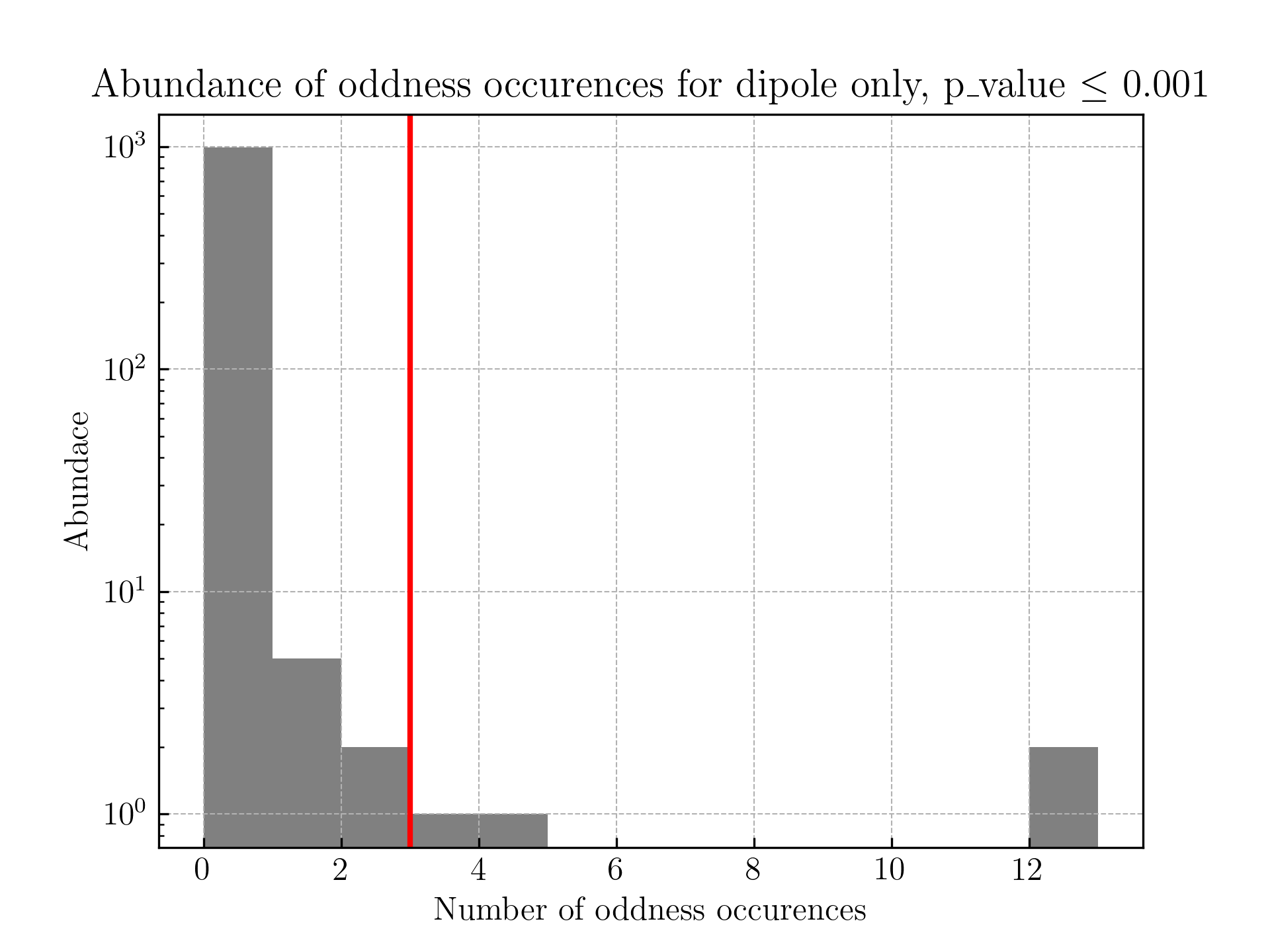}
	\includegraphics[width=9cm]{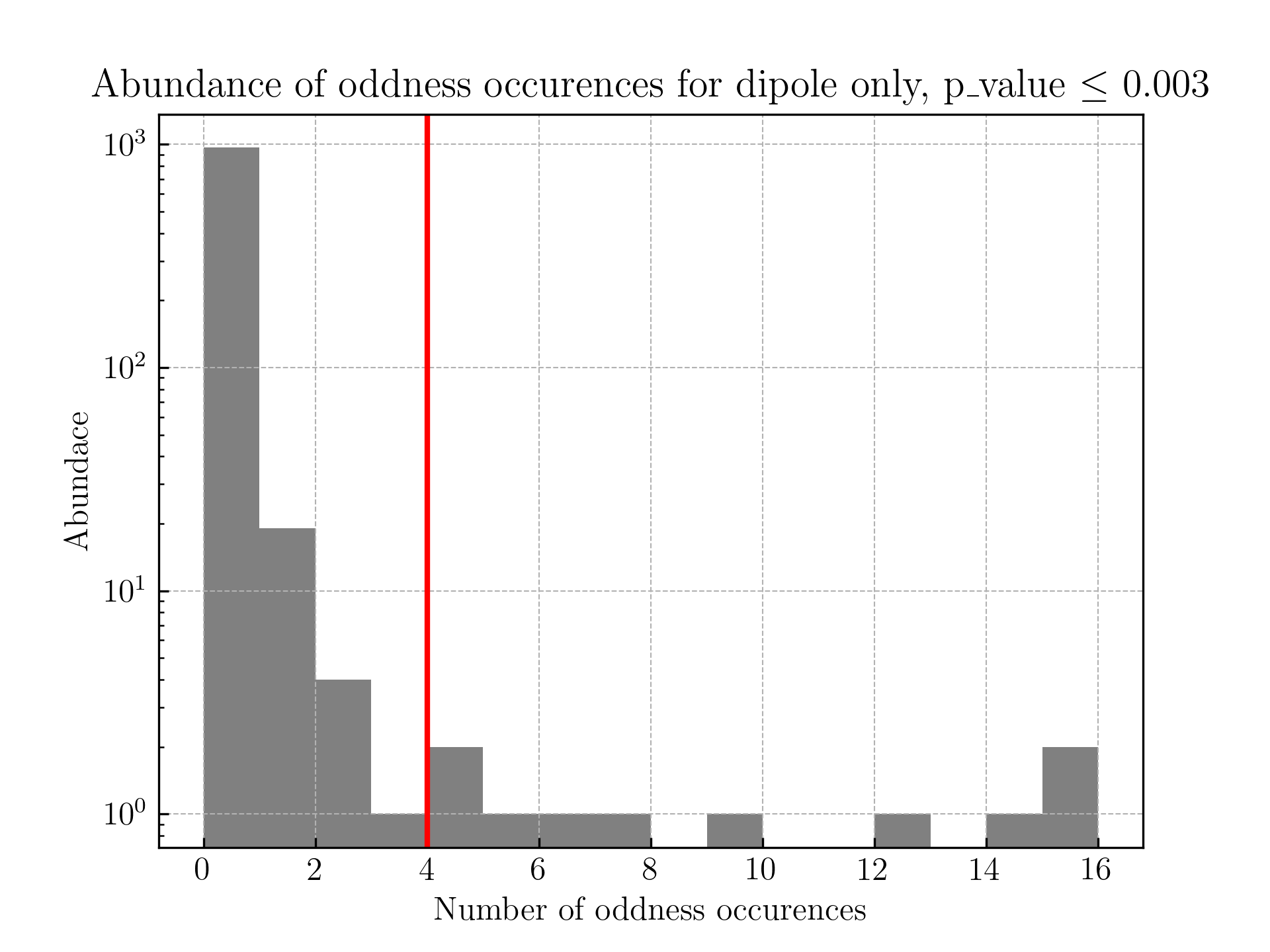}
	\caption{The abundance of simulations' oddness occurrences in $A_1$  parameter (i.e. dipole) for p-values of $\leq 0.001$ and $\leq 0.003$. This plot shows the significance of dipole anomaly and can be better understood if be compared to the next plot.}
	\label{dipole}
\end{center}
\end{figure}

\begin{figure}
	\begin{center}
		\includegraphics[width=9cm]{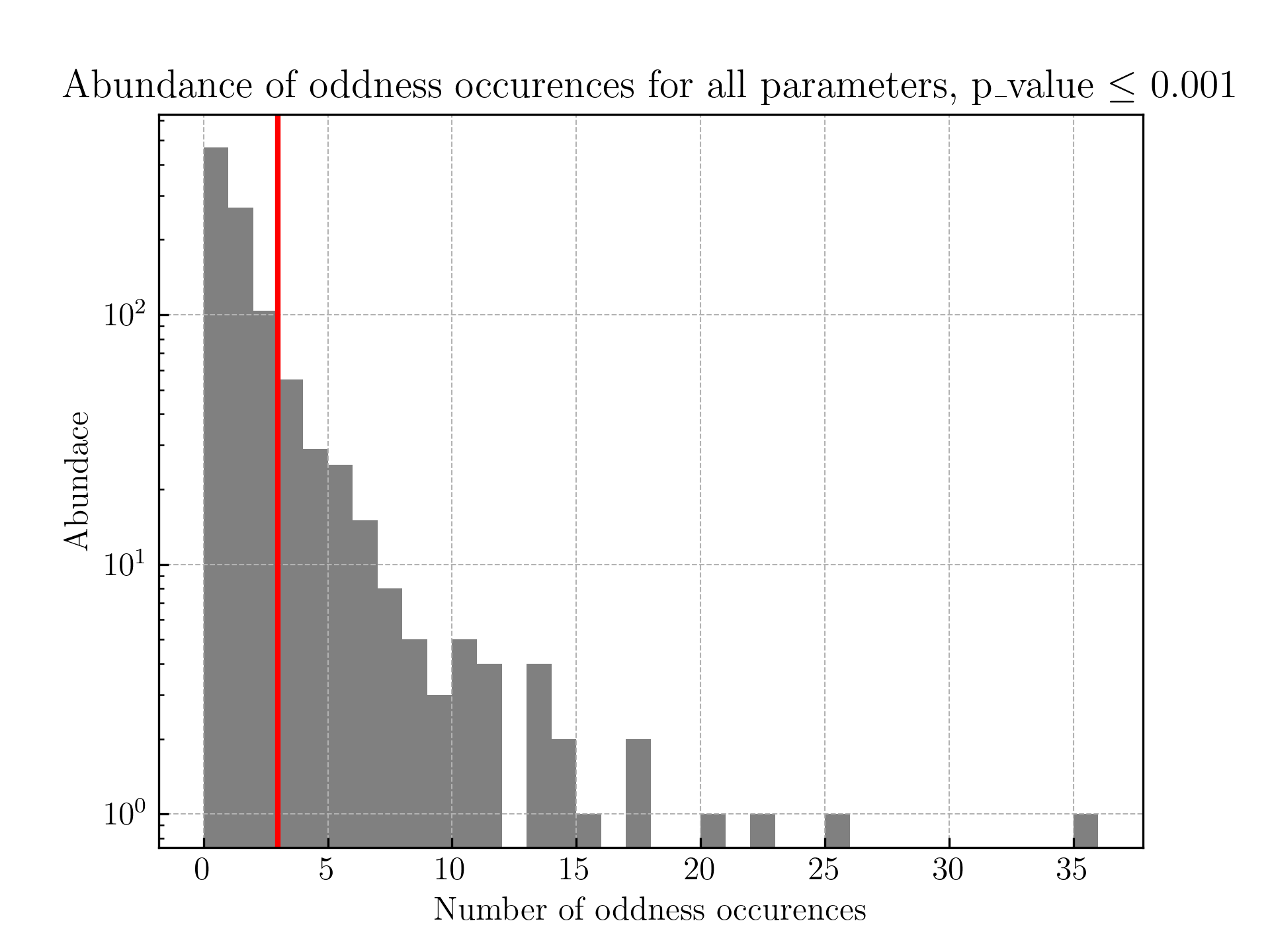}
		\includegraphics[width=9cm]{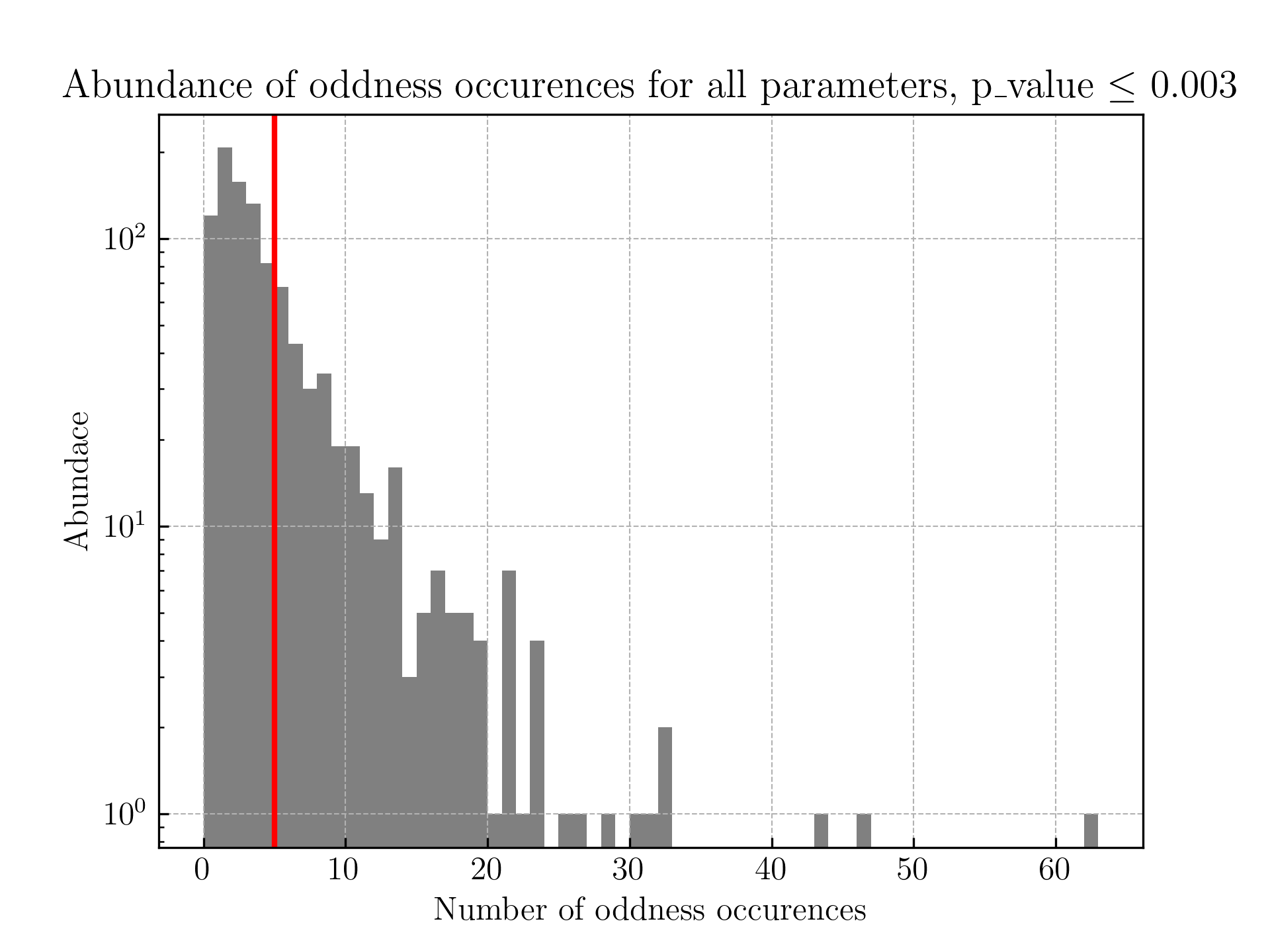}
		\caption{The abundance of simulations' oddness occurrences in all parameters for p-values of $\leq 0.001$ and $\leq 0.003$. The parameters mean different $A_\ell$'s for each map created with different cap sizes where for each simulation, there are about 600 parameters. The vertical red line indicates this measure for CMB. With this probe the CMB is not anomalous.}
		\label{all}
	\end{center}
\end{figure}

To conclude this section, and in order to gauge the statistical significance of higher multipoles outliers more rigorously, we aim to take into account the \textit{look-elsewhere-effect}.  This concern arises  because we have examined 600 quantities (20 disc sizes for multipoles up to $\ell=30$) and CMB being outlier in some of them might be expected just by chance. To have a look on this matter, we investigate the histogram of oddness occurrence for the simulation maps with two different thresholds: $p$-value $\leq0.001$ and  $p$-value $\leq0.003$. This means that we have counted number of all outlier quantities in every map (depending on the threshold of oddness) and then, based on these counted numbers, we deduce the distribution of oddness occurrence. If the CMB is an outlier in this distribution, one can conclude that it is indeed odd even when the look-elsewhere-effect  is compensated.  For instance, in Figure \ref{dipole} distribution of oddness is plotted, when only dipole ($\ell=1$) is considered with 20 different disc sizes. It is evident that CMB is outside of the distribution in this case. But when we consider all the multipoles up to $\ell=30$ in Figure \ref{all}, the CMB won't be an outlier anymore. However, this conclusion is an interpretation dependent result due to assuming that the dipole is more relevant in comparison to the higher multipoles. This is true when we start by assuming a (cosmologically) homogeneous and isotropic universe. In this universe, the dipole is the most relevant anisotropic feature in the cosmological scales (as the first term in the Legendre polynomials' series). However, we should emphasize that if e.g. there is a model in which the dipole anomaly is emerged by a small scale physics\footnote{As an instance, \citep{hansen2023possible} has reported an unknown effect from spiral galaxies on  photons that could be possible candidate for explaining several CMB anomalies} then we could expect the existence of a higher $\ell$ as relevant as the $\ell=1$. In this scenario we cannot trust to the above conclusion on look-eleswhere-effect.


\section{Concluding remarks and future perspectives}


In this work, we have analyzed the CMB variance maps searching for finer than a dipole structure. We have discussed that the previous statistical tools, like the angular power spectrum, that are based on statistical isotropy (except for the specific case of the dipole) were not proper tools for measuring such anomalies. We have assessed maps with different smoothing scales and found that several maps that show significant dipole amplitudes, also show anomalies in higher multipoles. Some multipole amplitudes are as significant as the dipole term, and in some cases, are even more significant than that. The analysis depends on the map preparation process (e.g. the smoothing scale) similar to the one looking for the dipole in the literature \citep{Planck:2019evm, Akrami:2014eta}. We think our analysis suggests more consideration of the isotropy tests of the sky, especially the CMB.

We could expect this feature from a theoretical viewpoint since having a pure observed dipole modulation in the CMB sky could mean that we are at the center of a dipole. This needs a fine-tuning in our place in the sky and theoretically disfavored. There are many models \citep{kanno2013viable,PhysRevLett111111302,DavidLyth2013,PhysRevD78123520,PhysRevD87123005,Gordon:2005ai,Abolhasani:2013vaa} in the literature, based on the physics of early universe,  tried to address the dipole anomaly. One different idea is studied in \citep{Koivisto:2007bp} where a dark energy model is responsible for this anomaly.  Our own dark energy model, GLTofDE which is constructed on the shoulders of critical phenomena's physics, could predict such a directional anomaly \citep{Banihashemi:2018has}\footnote{Actually, this feature in GLTofDE was our first motivation to look for the directioanl anomaly.}. The GLTofDE predicts the existence of patches with different sizes which produce a long wavelength mode. This long wavelength mode (via e.g. ISW effect) can produce an anisotropy in CMB. We think all of these theoretical explanations of dipole modulation anomaly have to be restudied and rechecked against the directional modulation anomaly.

In the observational and numerical side, we are in the first steps. In future works, we will try to optimize our numerical analysis by considering new probes and relaxing any symmetry in the map. If our analysis could be explained by a different patch in the CMB map then it would be great to check it with new CMB maps more accurately.

\section*{Acknowledgments}
We thank the anonymous referee for their beneficial suggestions and critiques related to our idea which makes our paper more reliable and informative. We are grateful to Yashar Akrami and Arman Shafieloo for very fruitful discussions on this project. We would like to thank MohammadHossein Jalali, Mahdieh Ebrahimi, and Bahar Torki for their help, in dealing with CMB maps. We want to express our gratitude to the Blender Foundation and community for providing the Blender software \citep{Blender}, which was instrumental in the completion of this work. The work of AB has been supported financially by Iran's Science Elites Federation.

\clearpage
\bibliography{power-asymmetry-refs}

\end{document}